\documentclass[aps,twocolumn,groupedaddress,showpacs]{revtex4}

\usepackage{graphicx}
\usepackage{amsmath}

\begin{document}

\title{Parton content of the real photon: astrophysical implications}

\author{I. Alikhanov}
\email[]{E-mail:ialspbu@gmail.com}

\affiliation{Institute for Nuclear Research of the Russian Academy of Sciences,
60-th October Anniversary pr. 7a, Moscow 117312, Russia}

\begin{abstract}
We possess convincing experimental evidence for the fact that the
real photon has non-trivial parton structure. On the other hand,
interactions of the cosmic microwave background photons with high energy particles propagating through the
Universe play an important role in astrophysics. In this context, to invoke the
parton content could be convenient for calculations of the
probabilities of different processes involving these photons. As an
example, the cross section of inclusive resonant $W^+$ boson production in the reaction
$\nu \gamma\rightarrow W^+X$ is calculated
by using the parton language. Neutrino--photon deep inelastic scattering is considered.
\end{abstract}


\maketitle


\section{Introduction}
The parton model, according to which hadrons consist of quarks
antiquarks and gluons (partons), bound together in different ways,
has been very successful in reproducing experiment. This provides a
relatively explicit and transparent technique for the description of
high energy particle interactions. The distributions of partons
inside hadrons are characterized by the structure functions
satisfying the Dokshitzer--Gribov--Lipatov--Altarelli--Parisi
(DGLAP) equations~\cite{lipatov,parisi,dokshitzer} or ones that are
basically similar. Such a function is the probability density of
finding a parton in a hadron carrying a fraction of the total
hadron's momentum. Numerical solutions of the equations are in a
remarkable agreement with experimental measurements, especially for
the nucleon~\cite{lai}.

Photons being involved in high energy interactions are also able to
manifest hadronic structure. One can intuitively comprehend this
since the photon directly couples to quarks and therefore may split
into quark--antiquark pairs. The parton contributions in two-photon
processes and some crucial peculiarities of the kinematic behavior
of the photon structure function have been described by Walsh and
Zerwas~\cite{zerwas}. The first work in studying
quantum-chromodynamics corrections to the naive pointlike structure
of the photon belongs to Witten \cite{witten}. This problem was also
studied in Refs.~\cite{dewitt,bardeen,duke}. Introducing the
evolution equations, similar to the DGLAP ones, for photons as well
as the properties of the corresponding solutions were under
scrutiny, for instance, in a series of papers by Gl\"uk, Reya,
Grassie and Vogt~\cite{gluk1,gluk2,gluk3}. A formulation of high
energy $\gamma p$ interactions taking into account the hadronic
properties of the photon was proposed in Ref.~\cite{schuler}.

Today, we possess convincing experimental evidence for the fact that the real photon has non-trivial parton structure \cite{cvach}.

On the other hand, the cosmic microwave background (CMB) photons may
play an important role in the formation of cosmic rays (CR). One of
the brightest representatives is the Greisen--Zatsepin--Kuzmin (GZK)
limit on the energy of CR~\cite{greisen,zatsepin}. For example,
protons of energies of over about $10^{20}$ eV would be decelerated
by interaction with the CMB photons, mostly due to resonant pion
production, $p\gamma\rightarrow\Delta^+\rightarrow p\pi^0(n\pi^+)$.
Other interesting processes, the $\nu\gamma$ reactions and their
possible astrophysical implications, were extensively discussed in
the literature (see,~\textit{e.g.},
Refs.~\cite{ex1,ex3,ex4,ex5,bugaev,ex8,rev1,rev2} and the references
cited therein). In this context, to invoke the parton content of the
real photon could be convenient for calculations of the
probabilities of such processes. Here, we attempt to show the
example of  $W^+$ boson production in the $\nu\gamma$ scattering
which may have important consequences for astrophysics~\cite{ex1}.
Studying this reaction could also provide a test of the universality
of the parton distribution functions of the photon.

\section{Neutrino--photon reactions within parton model}
Let us first consider inclusive on-shell $W^+$ boson production in
the reaction $\nu_e\gamma\rightarrow W^+X$ at the resonance region
using the parton language. We will view it from the center-of-mass
(CMS) frame of the $\nu_e\gamma$ system. Here, for example, a
substantial fraction of the CMB photons will be of energies of about
$(\varepsilon E^{lab}_{\nu})^{1/2}$, where $E^{lab}_{\nu}$ is the
neutrino energy in the laboratory frame defined as the frame in
which the CMB is isotropic, $\varepsilon$ is the CMB photon energy
(typical value $\varepsilon\sim10^{-3}$ eV~\cite{rev2}). This
reaction is standardly factorized into two subprocesses: the
emission of a positron by the photon and  annihilation of the
neutrino with the positron into $W^+$ (see Fig.~1a). Then the
corresponding cross section may be written as

\begin{figure}
\centering
\resizebox{0.4\textwidth}{!}{%
\includegraphics{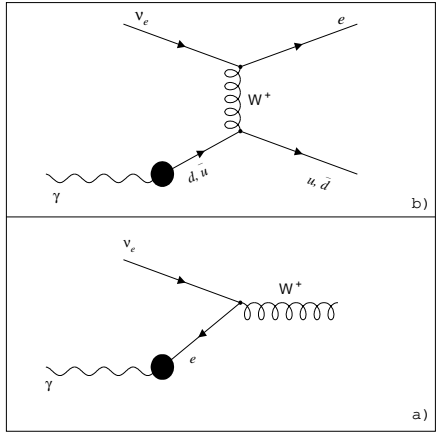}
}\caption{Diagrams illustrating \textbf{a)} the inclusive reaction
$\nu_e\gamma\rightarrow~W^+X$. Neutrino annihilates with positron
emitted by the photon into on-shell $W^+$; \textbf{b)} charged
current neutrino scattering off quarks (antiquarks) coming from the
photon. In this paper we take into account only the $u$ and $d$
quarks (antiquarks) and neglect Cabibbo--Kobayashi--Maskawa mixing.}
\label{fig1}
\end{figure}

\begin{equation}
\sigma(s)=\int_0^1\hat\sigma(xs)f_{\gamma}^{e}(x,s)\mathrm{d}x;
\label{eq:1}
\end{equation}

here $s$ is the total CMS energy squared ($s\simeq4\varepsilon
E^{lab}_{\nu}$), $f_{\gamma}^{e}(x,s)$ is the probability density
function to find the positron in the photon carrying the fraction
$x$ of the total photon's momentum, and $\hat\sigma(xs)$ is the
cross section of the annihilation subprocess. Note that we
explicitly write the $s$~dependence of the function instead of the
more traditional $Q^2$ one (4-momentum transfer squared) since we
deal with an $s$-channel subprocess.

In the resonance region $\hat\sigma(xs)$ is given by the Breit--Wigner formula~\cite{povh}

\begin{equation}
\hat\sigma(xs)=24\pi\frac{\mathrm{\Gamma_i}\mathrm{\Gamma}}{(xs-m_W^2)^2+m_W^2\mathrm{\Gamma}^2},
\label{eq:2}
\end{equation}

where $m_W$ is the mass of the $W^+$ boson, $\mathrm{\Gamma_i}$ is
the partial width of the initial channel (the partial width for the
decay $W^+\rightarrow\nu_e e^+$), and $\mathrm{\Gamma}$ is the total
decay width of $W^+$. In the leading order one can find that
\cite{lipatov2}

\begin{equation}
\mathrm{\Gamma_i}=\frac{G_Fm_W^3}{6\pi\sqrt{2}},\quad \mathrm{\Gamma}=9\mathrm{\Gamma_i},
\label{eq:3}
\end{equation}

where $G_F$ is Fermi's constant.

To determine the function $f_{\gamma}^{e}(x,s)$, we adopt the
formalism given in Ref.~\cite{gluk3}. It is fair to expect
$f_{\gamma}^{e}(x,s)$ to satisfy, up to factors associated with the
quark colors and fractional electric charges, the same evolution
equation as the quark distributions in the photon do, provided the
gluons are excluded and one takes into account only the
electromagnetic interaction. Then, in the leading order we write the
following equation for the positron distribution~\cite{gluk3}:

\begin{equation}
\frac{\mathrm{d}\,f_{\gamma}^{e}(x,Q^2)}{\mathrm{d}\ln Q^2}=\frac{\alpha}{2\pi}k^0(x),
\label{eq:4}
\end{equation}

where $\alpha$ is the fine structure constant,

$k^0(x)=2[x^2+(1-x)^2]$ \cite{furmanski,floratos}. Replacing in
Eq.~(\ref{eq:4}) $Q^2$ by $s$, for the reason explained above, and
choosing the electron mass squared $m_e^2$ as the lower integration
limit, one obtains

\begin{equation}
f_{\gamma}^{e}(x,s)=\frac{\alpha}{\pi}[x^2+(1-x)^2]\ln\frac{s}{m_e^2}.
\label{eq:5}
\end{equation}

Note that the latter result is similar, for example, to the one from Ref.~\cite{chen}.

Substituting Eqs.~(\ref{eq:2}) and~(\ref{eq:5}) into Eq.~(\ref{eq:1}) and performing the  integration, one finally arrives at the cross section

\begin{multline}
\sigma(s)=\frac{8}{3}\frac{\alpha\mathrm{\Gamma}^2}{s^3}\left[2s+\frac{s^2-2m_W^2(s+\mathrm{\Gamma}^2-m_W^2)}{\mathrm{\Gamma} m_W}\right.\\\times\left(\arctan{\frac{s-m_W^2}{\mathrm{\Gamma} m_W}}+\arctan{\frac{m_W}{\mathrm{\Gamma}}}\right)
\\
\left.+(s-2m_W^2)\ln{\frac{\mathrm{\Gamma}^2m_W^2+m_W^4}{\mathrm{\Gamma}^2m_W^2+(s-m_W^2)^2}}\right]\ln\frac{s}{m_e^2}.
\label{eq:6}
\end{multline}

The dependence of the cross section on $s$ is displayed in Fig.~2a
in comparison with calculations of the closely related process
$\nu_e\gamma\rightarrow W^+e^-$ carried out by Seckel~\cite{ex1}.
Here $m_W\simeq80.4$ GeV, $G_F\simeq1.16\times10^{-5}$ GeV$^{-2}$,
\newline $\alpha(m_W^2)\simeq1/128$~\cite{data}. One can see that
the values given by Eq.~(\ref{eq:6}) are about two times higher than
those of Ref.~\cite{ex1}.

\begin{figure}
\centering
\resizebox{0.45\textwidth}{!}{%
\includegraphics{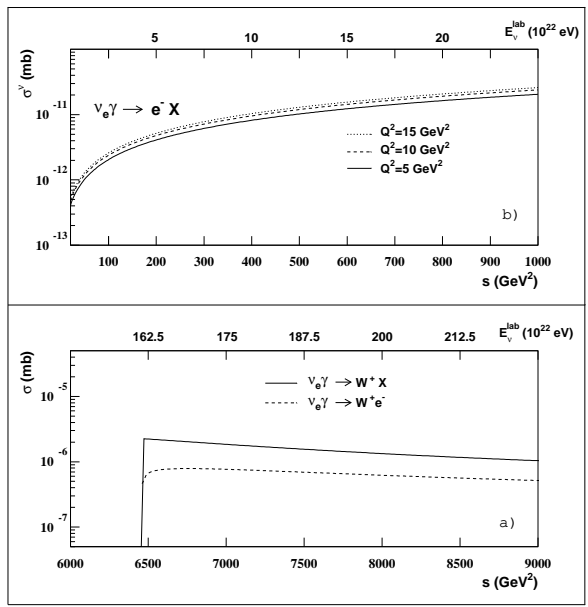}
}\caption{\textbf{a)} Dependence of the cross section of the
inclusive reaction $\nu_e\gamma\rightarrow W^+X$ on $s$ in the
resonance region (solid curve). The same calculated for the closely
related reaction $\nu_e\gamma\rightarrow~W^+e^-$~\cite{ex1} is shown
by the dashed curve. \textbf{b)} Dependence of the cross section of
the reaction $\nu_e\gamma\rightarrow~e^-X$ on $s$ at some fixed
values of $Q^2$ ($Q^2=~5$~GeV$^2$ -- solid curve, $Q^2=10$~GeV$^2$
-- dashed curve, $Q^2=15$~GeV$^2$ -- dotted curve). Note that the
laboratory energy of the neutrino $E^{lab}_{\nu}\simeq
s/4\varepsilon$ is calculated at $\varepsilon=10^{-3}$~eV.}
\label{fig2}
\end{figure}

Let us turn now to the charged current interaction of the neutrino
with the quark content of the photon (see Fig.~1b). The
corresponding cross section can be obtained in the same way as it is
done for neutrino--proton scattering~\cite{povh}:

\begin{equation}
\sigma^{\nu}(s)=\frac{G_F^2\,s}{\pi}\left(\frac{m_W^2}{m_W^2+Q^2}\right)^2\left(f_\gamma^q(Q^2)+\frac{1}{3}f_\gamma^{\bar q}(Q^2)\right),
\label{eq:7}
\end{equation}

with

\begin{equation}
f_\gamma^{q(\bar q)}(Q^2)=\int_0^1x\hat f_\gamma^{q(\bar q)}(x,Q^2)\mathrm{d}x,
\label{eq:8}
\end{equation}

where $\hat f_\gamma^{q(\bar q)}(x,Q^2)$ is the probability density
to find a quark $q$ (antiquark $\bar q$)  in the photon carrying the
fraction $x$ of the total photon's momentum. Taking into account
only the densities of the lightest quarks $u$ and $d$  from
Ref.~\cite{gluk2}, we found that

\begin{equation}
f_\gamma^{q}(Q^2)=e_{q}^2\frac{\alpha}{2\pi}\left(\ln\frac{Q^2}{m_{q}^2}-\frac{3}{4}\right);
\label{eq:9}
\end{equation}

here $e_q$ and $m_q$ are the electric charge and mass of the quark
$q$ respectively (for antiquarks the equation is analogous). Note
that Eq.~(\ref{eq:9}) is valid in the limit $m_q^2/Q^2\ll1$. We set
$m_u~=~m_d~=~0.2$~GeV and $\alpha=1/137$. The dependence of the
cross section thus determined on $s$ in the  range 20 GeV$^2$ $\leq
s\leq$ 1000 GeV$^2$ at some values of $Q^2$ is shown in Fig.~2b.

This reaction may have interesting astrophysical implications
because the struck quark may fragment into hadrons. The latter can
be highly boosted and on decaying (if unstable) may produce
particles with energies exceeding their GZK limit. If it occurs in
the vicinity of the Earth the decay products may reach us without
significant energy loss, provided the incident quark momentum
pointed in the direction of the Earth. A similar idea has been
proposed, for example, in Ref.~\cite{weiler}, when photons would
appear beyond the GZK limit from decays of highly boosted  $\pi^0$,
which, in turn, were the decay products of real $Z^0$ bosons excited
in $\nu\bar\nu$ annihilation (the so-called "$Z$-burst" mechanism).
But there are problems here, mainly associated with the origin of
such high energy neutrinos, $E^{lab}_{\nu}\simeq m_Z^2/4\varepsilon$
(see, \textit{e.g.}, Ref.~\cite{bugaev}). In our case, the minimal
neutrino energy required to produce hadrons is smaller than the
latter one by about a factor of 400, and the corresponding cross
section is also suppressed by a factor $\alpha G_F$. Anyway, one
may expect that these processes were important for high energy
neutrino absorption in the early Universe.

Throughout this paper we implicitly  used the assumption that the
parton distributions are process--independent, which has been
experimentally justified for the nucleon. For example, the functions
phenomenologically derived from electron--nucleon and
neutrino--nucleon deep inelastic scattering data are close to each
other. Using them one can correctly predict the probabilities of
inclusive production of $\mu^+\mu^-$~pairs in $p\bar p$~collisions
(Drell--Yan process) \cite{close}.

Analogously, the neutrino--photon reactions could provide an
instrument for studying the universality of the parton distributions
in the photon.

We have discussed only the $\nu_e\gamma$ interactions. Meanwhile,
all the things we said above may be straightforwardly applied to the
reactions involving  the antineutrino. Likewise, heavier charged
leptons can be considered. One may also include neutral current
interactions in the neutrino--quark scattering. Other processes
involving the CMB photons can be treated in similar way.


\end{document}